\documentclass[aps, twocolumn]{revtex4-1}
\usepackage{graphicx}
\usepackage{dcolumn}
\usepackage[utf8]{inputenc}
\usepackage[T1]{fontenc}
\usepackage{bm}
\usepackage{ulem}
\usepackage{mathptmx}
\usepackage{amssymb}
\usepackage{etoolbox}
\usepackage{booktabs}
\usepackage{amsmath, amsfonts}
\usepackage{tikz}
\usepackage{enumerate}
\usepackage{url}
\usepackage{float}
\usepackage{wasysym}
\usepackage[mathscr]{eucal}
\usepackage{hyperref}
\usepackage{svg}
\usepackage{wrapfig}

\hypersetup{
    colorlinks=true,
    linkcolor=blue,
    filecolor=magenta,      
    urlcolor=cyan,
    }

\usetikzlibrary{shapes.geometric}
\usetikzlibrary{decorations.pathreplacing, angles, quotes}
\usetikzlibrary{arrows, decorations.markings}
\usetikzlibrary{plotmarks}
\usepackage{rotating,array}
\usepackage{xcolor}
\usepackage[utf8]{inputenc}
\definecolor{MATLABblue}{HTML}{0072BD}
\definecolor{MATLABorange}{HTML}{EDB120}
\definecolor{CTD}{rgb}{0.25,0.65,0.25}

\begin{document}
\title{Boltzmann-Shannon Index: A Measure of Density Balance in Clustered Data}
\author{Emanuele Bossi} 
\affiliation{Embry-Riddle Aeronautical University, Prescott, AZ 86301}

\author{C. Tyler Diggans}
\affiliation{Root Dynamix LLC, Sewickley, PA 15143}

\author{Abd AlRahman R. AlMomani} 
\affiliation{Embry-Riddle Aeronautical University, Prescott, AZ 86301\relax}

\begin{abstract}
    % App. 100-words abstract. If a longer abstract is needed, I can re-draft this.
    The Boltzmann–Shannon Index (BSI) for clustered continuous data is introduced as a normalized measure that captures the relationship between geometry-based and frequency-based probability distributions defined over the clusters. In essence, it quantifies the similarity across densities of the clusters, which are defined by a given labeling.  This labeling may originate from a geometric partitioning of the state space itself, but need not in general.  We illustrate its performance on synthetic Gaussian mixtures, the Iris benchmark data set, and a high-imbalance resource-allocation scenario, showing that the BSI provides a coherent assessment in cases where traditional metrics give incomplete or misleading signals. Moreover, in the resource-allocation setting where equal density may be associated with a ``fair'' distribution, we demonstrate that BSI not only detects inequality with high sensitivity, but also offers a numerically smooth measure that can be easily embedded in optimization frameworks as a regularization term for modern policy-making.  Finally, the BSI also offers a new measure of the effectiveness for a given symbolic representation, i.e. coarse-grain states, for continuous-valued data recorded from complex dynamical systems.
\end{abstract}

\maketitle

\section{Introduction}
Entropy, as a measure of disorder, was the cornerstone of the foundation of statistical mechanics.  Although in that context it was defined in terms of volumes of phase space that contain a density of microscopic configurations of a given macrostate for a thermodynamic system\nobreakspace\cite{boltzmann1866mechanische}, it makes clear that ``the logarithmic connection between entropy and probability was first stated by Ludwig Boltzmann in his kinetic theory of gases''\nobreakspace\cite{planck1914theory}.  

It was almost a century later that Claude Shannon re-framed the concept of entropy in the more abstract context of information theory, where he defined it as a quantitative measure of uncertainty over a discrete probability distribution used in symbolic communication.  He established a framework using this concept for understanding the fundamental limits of signal processing and data compression, with a focus on the concept of channel capacity\nobreakspace\cite{shannon1948mathematical}.  And, while Shannon’s formulation is similarly a cornerstone of modern data science, it was designed for applications in discrete symbolic state spaces.  

All applications to continuous state spaces, therefore, involve specific assumptions about smooth probability mass functions (outlined in the final chapter of his seminal work\nobreakspace\cite{shannon1948mathematical}).  However, as E. T. Jaynes emphasized in his reformulation of information theory through a more Bayesian lens\nobreakspace\cite{jaynes1957information}, these assumptions are not self-evident. Most importantly, Shannon entropy is always defined relative to an underlying measure, and so any approach based on standard Riemann sums would correspond to choosing a uniform prior over the bins, i.e. arbitrarily chosen states. Or as Jaynes put it:   
\begin{quote}
    "For on the frequentist view, the notion of a probability for a person with a certain state of knowledge simply doesn't exist, because probability is thought to be a real physical phenomenon which exists independently of human information. But the problem of choosing some probability distribution to represent the information source still does exist; it cannot be evaded. It is now clear that the whole content of the theory depends on how we do this." \cite{jaynes1995probability}
\end{quote}

When analyzing data in continuous state spaces, frequency of observation is not the only measure of interest; another important factor to consider is the geometry of the distribution of data.  Data points occupy space, form shapes, and concentrate in regions of varying density. For this reason, recent work explored a specific form of metric entropy\nobreakspace\cite{sinai1959notion,kolmogorov1985new}, termed \textit{geometric partition entropy}\nobreakspace\cite{diggans2022geometric, diggans2025generalizing}, which defines entropy directly from the spatial organization of data rather than from symbolic frequencies. Instead of using traditional histogram binning approaches, Geometric Partition Entropy (GPE) partitions the state space into coarse-grained regions so that each bin contains approximately the same number of sample points (e.g., using quantiles in one dimension).  The entropy is then estimated from the distribution of proportions each state takes up within a bounded state space, providing a geometry-driven measure of uncertainty.  This is akin to a Lebesgue integral formulation where the range of the CDF, $[0,1]$, is binned, as opposed to the domain in the Riemann integral of Shannon (see Fig.\nobreakspace\ref{fig:DSEGPE}, reproduced from \cite{diggans2023boltzmann} with the permission of the authors, for an illustration of the relationship between a GPE estimate and the corresponding Riemann estimate of Shannon entropy).

\begin{figure}[ht!]
    \centering
    \includegraphics[width=0.77\linewidth]{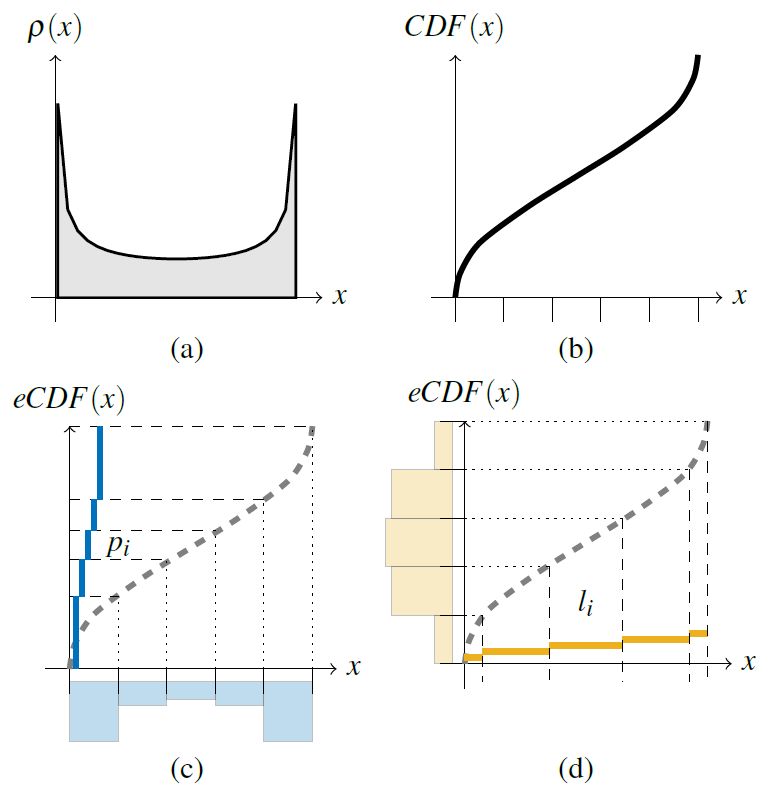}
    \caption{\textit{Comparison between frequency-based and geometry-based quantifications of uncertainty in one-dimensional data}. An example of (a) an underlying probability distribution $\rho(x)$ together with (b) a corresponding cumulative distribution function $CDF(x)$. A traditional Riemann sum (frequency-based) estimate for entropy uses (c) a distribution defined by the differences in the empirical CDF at fixed equal length bin boundaries; the probability $p_i$ of the $i$-th bin is given by the height of the corresponding blue step. In contrast, geometric partition entropy (GPE) constructs a measure-based representation as in (d) by partitioning the co-domain $[0,1]$ into equal masses and mapping the boundaries back through the inverse CDF; the probability $q_i$ is obtained from the geometric length $l_i$ taken as a proportion of the total state-space length. This figure was reproduced with the permission from the authors of \cite{diggans2023boltzmann}.\label{fig:DSEGPE}}
\end{figure}

While GPE can provide more stable estimates of entropy for data in continuous state spaces, especially for small sample sizes, this geometric approach is not always more informative or robust. For example, when samples contain repeated or nearly identical values, the quantile-defined bins can collapse to zero measure, yielding degenerate partitions and uninformative entropy estimates\nobreakspace\cite{diggans2023boltzmann}. In these kinds of scenarios, GPE fails to distinguish genuine geometric concentration from mere repetition.

Thus, for many complex data distributions there is clear utility in both viewpoints.  In fact, many advances in information theory, e.g., the popular KNN method for estimating mutual information, have originated from the inclusion of geometry and frequency.  This trend perhaps started with the work of Kozachenko and Leonenko\nobreakspace\cite{kozachenko1987sample}, and the \textit{Boltzmann-Shannon Interaction Entropy} (BSIE)\nobreakspace\cite{diggans2023boltzmann} sought to highlight that earlier work.  By combining the Boltzmann-inspired geometric view of GPE with Shannon’s frequentist formulation, BSIE provided a normalized measure that captured how the underlying geometry of a set of observations of continuous variables interacts with a corresponding estimated probability mass function. Like that earlier work, rather than treating shape and frequency as separate aspects of a dataset or averaging away its effects (as in KNN), BSIE sought to quantify the joint structure of a pair of related distributions (e.g., see \label{fig:DSEGPE}).

While BSIE provided a measure that incorporated both the geometry and frequency of data in continuous variables, it did not directly account for scenarios in which the data are already partitioned (i.e., clustered), which is a common occurrence.  In this case, labels are assigned that group the data into a number of disjoint subsets.  The challenge is then not necessarily to quantify the entropy of a distribution, but rather to understand the impact a given choice of partition has on the resulting symbolic representation through the underlying geometry and density of the states.

For example, it is common practice to label a dataset as imbalanced solely on the basis of the target distribution, typically when one class accounts for the vast majority of observations. But, when data lie in continuous spaces, imbalance may not be determined by frequency-based probability alone. For example, consider a scenario where the outcome variable follows a $90:10$ distribution, but the underlying state space is also partitioned in a way that reflects the same $90:10$ proportion. In such a case, the dataset should not be considered imbalanced.  The frequency distribution is fully consistent with the geometric structure of the data, and the representation of data proportionally expresses the regions of interest in the feature space. True balance or imbalance therefore arises from the interaction between class frequencies and the geometry of the feature space, not from frequencies in isolation.

Here, we introduce the \textit{Boltzmann-Shannon Index} (BSI), which gives a sense of how well the state space is partitioned into clusters where equal density clusters would result in unity.

Unlike existing clustering validity indices, the proposed BSI captures both the geometric spread of the data and the frequency with which regions of the state space are occupied. Distance-based metrics, such as the Silhouette Score \cite{rousseeuw1987silhouettes}, Davies–Bouldin Index \cite{davies2009cluster}, and Calinski–Harabasz Index \cite{calinski1974dendrite}, quantify compactness or separation but rely solely on pairwise distances or variance ratios, ignoring how density is distributed across clusters.  Moreover, they are not normalized and so can be difficult to compare across datasets or scales. Information-theoretic measures, such as normalized mutual information \cite{strehl2002cluster}, overcome some of these issues but require ground-truth labels and therefore cannot be used in fully unsupervised settings. In contrast, the Boltzmann–Shannon Index is agnostic, normalized, and uniquely incorporates both geometry-based and frequency-based probabilities, enabling it to detect imbalances in how the state space is partitioned that traditional metrics cannot reveal.

We define the Boltzmann-Shannon Index in Sec.\nobreakspace\ref{sec:Methodology}, and apply this measure to several data sets with increasing relevance and complexity in Sec. \nobreakspace\ref{sec:Results}.  We then conclude with a summary of the work and plans for adaptation to mixed continuous/discrete datasets and offer several directions for future applications.

\section{Methodology} \label{sec:Methodology}
We define the Boltzmann-Shannon Index for a dataset together with labels that partition the observations into $K$ disjoint subsets.  Although these labeled subsets often correspond to a natural geometric partition of the state space, this need not be the case.  In other words, the regions associated with different labels may overlap geometrically, provided that each data point is assigned exactly one label. 

For the purposes of this work, the specific algorithm used for clustering is irrelevant as the BSI operates "\textit{post-hoc}" on any set of cluster labels. Its role is to quantify how effectively a given clustering or partitioning scheme organizes a continuous state space into a meaningful coarse-grained representation.  Since our interest lies in evaluating a given partition rather than in how it is produced, we need only rely on the common $K$-Means clustering\nobreakspace\cite{lloyd1982least}, which partitions data into $K$ clusters by minimizing within-cluster variance.

Given a data set $\mathcal{X}=\left\{x_0,x_1,...,x_N\right\}\subset \mathbb{R}^d$ and labels $\mathcal{L}=\left\{\ell_0,\ell_1,...,\ell_N\right\}$ where $\ell_i\in\left\{1,...,K\right\}$, we define the Boltzmann-Shannon Index to be one minus the Jensen Shannon Divergence (JSD) between a geometric measure-based distribution over the $K$ states of the partition, i.e. clusters, and the frequency-based distribution of the same $K$ states.  If we let $\textbf{q}$ represent a normalized geometric distribution and $\textbf{p}$ represent the commonly used normalized frequency-based distribution (i.e. a histogram), then we define the BSI as:
\begin{equation}
\begin{aligned}
BSI(\mathcal{X};\mathcal{L}) &= 1-JSD(\textbf{p}||\textbf{q})\\
& = 1-\frac{D_{KL}(\textbf{p}||\textbf{m})+D_{KL}(\textbf{q}||\textbf{m})}{2},
\end{aligned}
\end{equation}
where $D_{KL}(x||y)$ is the Kullback-Leibler divergence between distributions $x$ and $y$, and $\textbf{m}$ is taken to be the midpoint distribution defined as the average of $\textbf{p}$ and $\textbf{q}$, i.e. $\textbf{m}=(\textbf{p}+\textbf{q})/2$.  

Although the Kullback-Leibler divergence is asymmetric, the use of the midpoint distribution in the JSD results in a symmetric and normalized measure. When the distributions $\textbf{p}$ and $\textbf{q}$ are essentially the same over the $K$ states, we end up with a very small value for the $JSD(\textbf{p}||\textbf{q})$ and thus we get a value very close to one for the BSI.  At the other extreme, when the geometries of the $K$ states and their frequencies are poorly aligned, reflecting substantial density imbalances across states, the $JSD(\textbf{p}||\textbf{q})$ increases toward one, yielding a correspondingly small BSI value.  

It is important to note that, unlike the Boltzmann-Shannon Interaction Entropy ––– where two distinct partitions are used to define $\textbf{p}$ and $\textbf{q}$ ––– the BSI uses the same categorical bins that are determined by the given clustering or labeling. The two distributions $\textbf{p}$ and $\textbf{q}$ are instead constructed using different underlying measures, one frequency-based and one geometry-based.  Further, while the frequency-based distribution is common and self-explanatory, the geometric measure-based distribution may be defined or estimated in a number of meaningful ways. However, it requires that clusters corresponds to finite subsets of $\mathcal{X}$ within a bounded domain $\mathcal{D}\subset\mathbb{R}^d$ for some $d\in\mathbb{N}$, ensuring that their geometric measures can be properly normalized. 

The most obvious measures arise in the case of geometrically separable clusters that allow a true partition of $\mathcal{D}$ into $K$ disjoint regions, $R_1,R_2,...,R_K$, such that their union equals $\mathcal{D}$, e.g. based on a Voronoi diagram.  Here, $q_i$ may simply be the $d$-dimensional volume of the region $R_i$.  However, it is very common to have data where no clear bounded domain is known \textit{a priori}; in that case, there are many approaches that could be taken.  This challenge was addressed in detail in previous work\nobreakspace\cite{diggans2022geometric,diggans2023boltzmann} and so we will not revisit it here; instead, we simply suggest a measure that works in all cases including the partition described above.  

Since we do not want to enforce geometric separability of the clusters by label, we will rely on the powerful connection of \textit{Singular Value Decomposition} (SVD) to geometry. Given labeled data, as stated above, we consider the rectangular matrices of each labeled cluster one at a time, where rows of each matrix are the $d$-dimensional vectors in a given labeled subset.  We compute the SVD of each subset and multiply the $d$ singular values to get a geometric quantity that is the product of the variation in each of the principal component direction.  We chose this as our suggested geometric measure because it is defined for all partitions, even when concepts like the convex hull of the subsets might overlap and be skewed by outliers.

Coarse-grained states representing real dynamical time series data are rarely equal in their geometric extent or frequency of occurrence. One goal of the proposed index is to provide a normalized measure of how close such a clustering of data is to an equal density distribution, which often results in more expressive transition matrices.  The Boltzmann–Shannon Index then offers an immediate, parameter-free diagnostic of the degree to which a given partition achieves balanced density across clusters, a property that is highly desirable in exploratory data analysis, yet invisible to conventional validity indices.

\section{Results} \label{sec:Results}
To illustrate the behavior and interpretability of the Boltzmann–Shannon Index, we evaluate its performance across a sequence of increasingly realistic scenarios, beginning with simple synthetic data and progressing toward more application appropriate resource-allocation problems.

\subsection{Two-cluster example: frequency versus geometry reversal}
\label{sec:two-cluster}

To provide intuition for the Boltzmann–Shannon Index, we begin with the simplest non-trivial setting: a data set partitioned into exactly two clusters. Let the empirical frequency distribution be:
\begin{equation}
    \textbf{p} = (\,p_1,\; p_2\,) \;=\; (\alpha,\; 1-\alpha),\qquad \alpha\in[0,1].
\end{equation}

Here, the first cluster contains the proportion $\alpha$ of the points, and the second one the remaining $1-\alpha$. For the geometric distribution $\textbf{q}$, we deliberately choose the \textit{reversed} ordering:
\begin{equation}
    \textbf{q} = (\,q_1,\; q_2\,) \;=\; (1-\alpha,\; \alpha).
\end{equation}

This corresponds, for example, to the situation in which the larger cluster (in population) occupies the smaller volume in feature space, and the smaller cluster is spread over a much larger volume; a classic form of severe density imbalance. Such patterns frequently arise in real-world data, for instance in anomaly-detection settings where rare events (e.g., network intrusions or equipment faults) occupy broad, diffuse regions of the state space while normal behavior is highly concentrated. The Boltzmann-Shannon Index for $\textbf{p}$ and $\textbf{q}$ is then defined analytically as:
\begin{eqnarray}
\operatorname{BSI}(\alpha) 
&=& 1 - \operatorname{JSD}(\textbf{p}\Vert \textbf{q}) \notag \\
&=& 1 - \frac{1}{2}\Bigl[D_{\mathrm{KL}}\bigl(\textbf{p}\Vert \textbf{m}\bigr) 
                     + D_{\mathrm{KL}}\bigl(\textbf{q}\Vert \textbf{m}\bigr)\Bigr] \notag \\
&=& 1 - \frac{1}{2}\Biggl[\alpha\log_2\!\Bigl(\frac{\alpha}{1/2}\Bigr) 
                         + (1-\alpha)\log_2\!\Bigl(\frac{1-\alpha}{1/2}\Bigr) \notag \\
&=& 1 - \frac{1}{2}\Bigl[2\alpha(\log_2\alpha + 1) 
                       + 2(1-\alpha)\log_2(1-\alpha) + 2\Bigr] \notag \\
&=& 1 - \Bigl[\alpha\log_2\alpha + (1-\alpha)\log_2(1-\alpha) + 1\Bigr] \notag \\
&=& -\alpha\log_2\alpha - (1-\alpha)\log_2(1-\alpha).
\label{eq:two-cluster-derivation}
\end{eqnarray}
Thus, under this deliberate frequency-geometry reversal, the Boltzmann-Shannon Index reduces precisely to the entropy of a Bernoulli random variable with success probability $\alpha$, or, equivalently, to the mutual information between the cluster label and an ideal uniform geometric prior.
\begin{figure}[ht]
\centering
\includegraphics[scale=0.8]{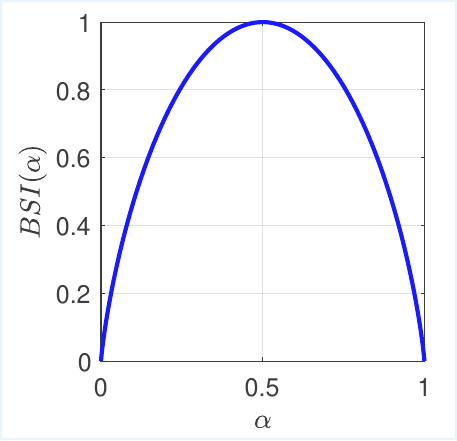}
\caption{\textit{Two-cluster reversal example}. Population frequency $\textbf{p} = (\alpha,1-\alpha)$ and geometric spread $\textbf{q} = (1-\alpha,\alpha)$. 
The Boltzmann–Shannon Index (solid blue curve) attains its maximum value of 1 at $\alpha = 0.5$, corresponding to perfect frequency–geometry alignment, and decreases towards 0 as $\alpha\to 0^+$ (or $\alpha\to 1^-$), where the two distributions are completely inverted and nearly all points occupy the geometrically smaller (or larger) region.}
\label{fig:two-cluster}
\end{figure}

As showed in Fig.\nobreakspace\ref{fig:two-cluster}, $\operatorname{BSI}(\alpha)$ attains its theoretical maximum of one when population and geometry are perfectly matched ($\alpha = 0.5$); conversely, when population and geometry are perfectly inverted (in either case $\alpha \to 0$ or $\alpha \to 1$), the index correctly approaches 0, signaling extreme density imbalance. This elementary two-cluster example already reveals the core contribution of the Boltzmann–Shannon Index in that it rewards partitions in which cluster sizes are proportional to the geometric volume (or effective spread) they occupy, and penalizes the opposite situation.

The remaining experiments in this section will show that the same intuitive pattern persists across synthetic mixtures, benchmark datasets and socio-economic resource-allocation problems, but additional interesting details emerge.

\subsection{Synthetic Gaussian mixtures}

\begin{figure}[ht!]
\centering
\begin{tabular}{cc}
\includegraphics[scale=0.6]{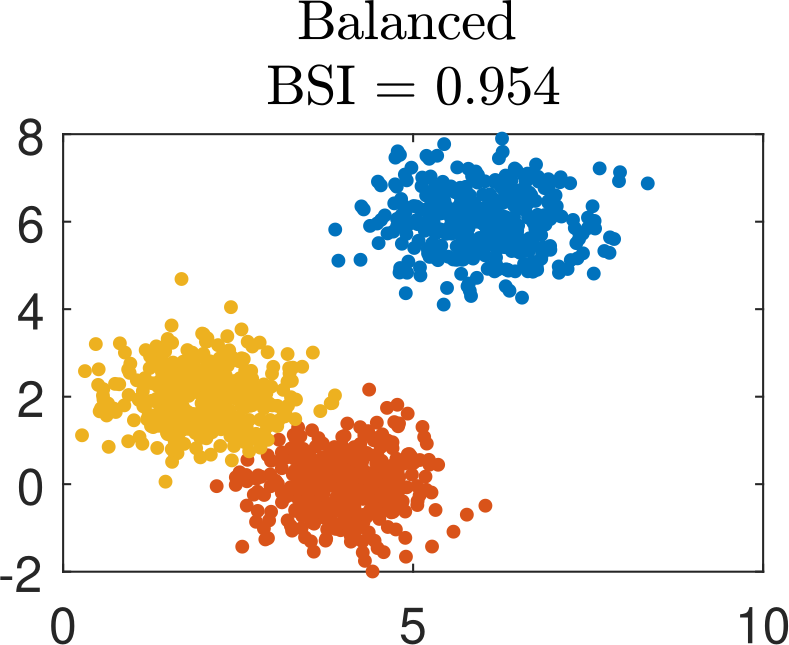}& \includegraphics[scale=0.6]{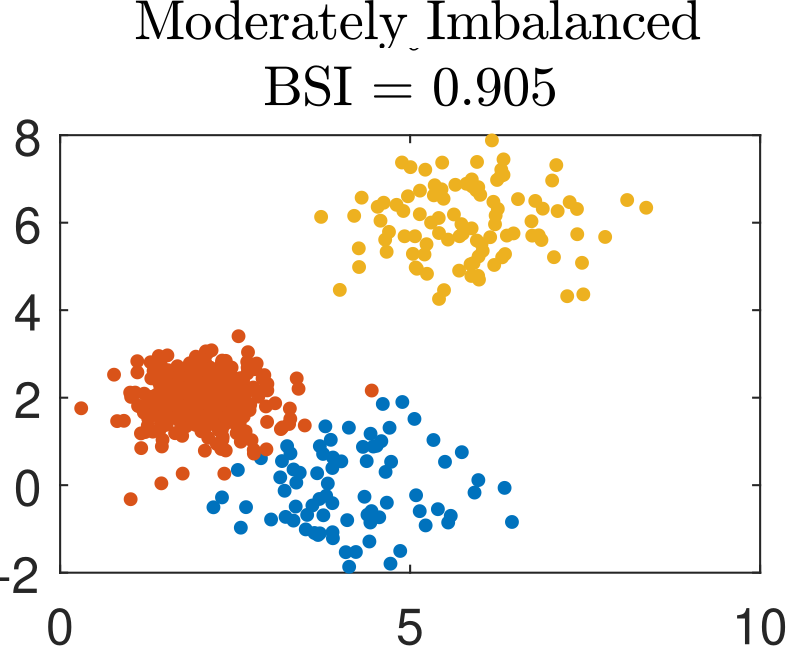}\\
(a) & (b)\\
\end{tabular}
\includegraphics[scale=0.75]{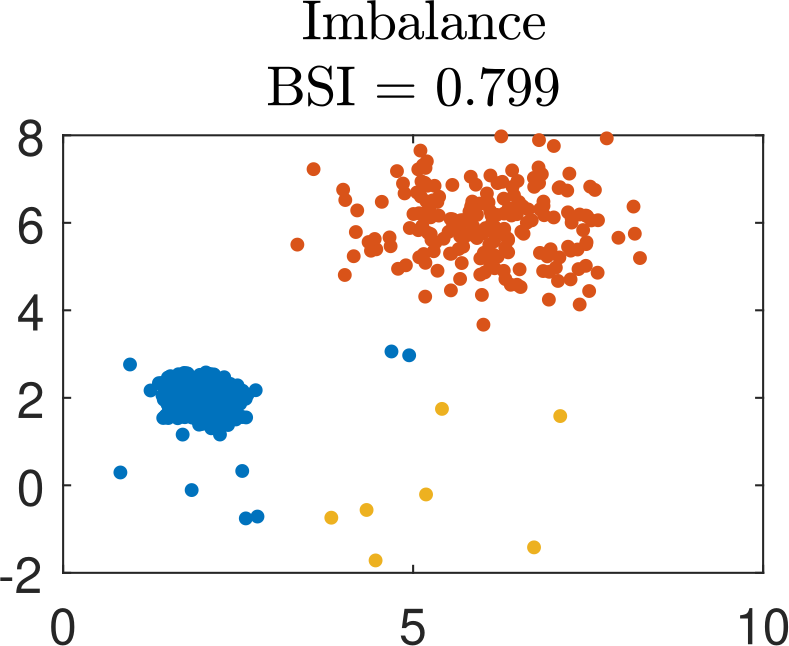}\\
(c) \\
\caption{\textit{Synthetic 2D Gaussian mixtures illustrating the behavior of the Boltzmann–Shannon Index}. 
(a) balanced and well-separated clusters (high BSI); (b) moderately imbalanced sizes (intermediate BSI); and 
(c) strongly imbalanced and overlapping clusters (low BSI).\label{fig:gaussians}}
\end{figure}

Fig.\nobreakspace\ref{fig:gaussians} includes three deliberately constructed two-dimensional data sets (a)-(c), each generated from a mixture of three isotropic Gaussians. The first mixture (a) is balanced and well-separated, with comparable cluster sizes and nearly identical variances; the second (b) introduces a moderate size imbalance while maintaining nearly identical variances; and the third (c) combines a strong imbalance in population with substantial overlap, yielding a visibly degenerate and challenging partition.

In each case, we applied standard $K$-Means clustering with $K=3$, using random initialization and multiple restarts (empty-cluster action = singleton). The resulting labels were then passed directly to the Boltzmann–Shannon Index, which employs the SVD-based geometric measure $\textbf{q}$ described in Sec.\nobreakspace\ref{sec:Methodology}.

This example shows that the BSI reliably captures the degree to which cluster membership is aligned with the geometric spread each cluster occupies in the feature space, which is an intuitively desirable property that cannot be recovered by frequency-only validity measures.

\subsection{Iris benchmark dataset}

We also apply the Boltzmann-Shannon Index to the classic Iris dataset\nobreakspace\cite{fisher1936use}, which contains $150$ observations with four continuous-valued features describing the three species of \textit{setosa}, \textit{versicolor}, and \textit{virginica}, each represented by exactly $50$ samples, an objectively near equal density case.

Although the ground-truth species labels are known, we deliberately ignore them during clustering and performed standard $K$-means clustering with $K=3$, using $200$ independent restarts and the singleton remedy for empty clusters. The best run, i.e. the one maximizing the BSI, resulted in a partition that perfectly separates \textit{setosa} and merges only a few \textit{versicolor-virginica} points, yielding the values reported in Table \ref{tab:iris}.

\begin{table}[ht!]
\centering
\caption{Clustering quality indices on the Iris dataset ($K$-means, $K=3$, best of 200 restarts).}
\begin{tabular}{lc}
\toprule
Metric                                    & Value          \\
\midrule
Boltzmann-Shannon Index (proposed)       & \textbf{0.9901} \\
Boltzmann-Shannon Index (ground-truth labels) & 0.9952     \\
Silhouette score                          & 0.7357         \\
Calinski-Harabasz index                  & 561.6          \\
Davies-Bouldin index                     & 0.6620         \\
Shannon entropy of cluster sizes (bits)   & 1.5570         \\
\bottomrule
\end{tabular}
\label{tab:iris}
\end{table}

The Boltzmann-Shannon Index applied to the best result from $K$-means reaches $0.990$, a value extremely close to the theoretical maximum of $1.0$ and only marginally below the value of $0.9952$ that is obtained when using the true species labels. This near-unity score reflects the fact that the recovered clusters are not only of almost perfectly equal cardinality but also occupy nearly identical effective volumes in the four-dimensional feature space, as measured by the SVD-based geometric distribution $\textbf{q}$.

Traditional indices provide a noticeably less coherent assessment. The silhouette score of $0.74$ is reasonable but substantially lower than the BSI, as it reflects only local separation and is unavoidably penalized by the well-known overlap between \textit{versicolor} and \textit{virginica}. The Calinski-Harabasz and Davies-Bouldin indices yield favorable values, yet both are driven primarily by centroid distances and within-cluster dispersion, making them largely insensitive to the striking balance in cluster occupancy. Even the ordinary Shannon entropy of the cluster sizes is already close to its maximum possible value of $\log_2 3 \approx 1.585$ bits, and would remain high even if one cluster collapsed to only a few scattered outliers, highlighting its inability to detect geometric degeneracy or mismatches between population and spatial extent.

In contrast, the Boltzmann-Shannon Index simultaneously verifies that (i) the clusters are evenly populated and (ii) each cluster occupies a comparable volume in the feature space. Because the Iris dataset exhibits this ideal alignment between frequency and geometry, the BSI approaches unity. Crucially, even slight perturbations that introduce either population imbalance or geometric distortion, such as compressing one cluster or spreading another, would cause a noticeable drop in BSI well before traditional metrics register a significant change. This sensitivity confirms that the proposed index is a stricter and more complete diagnostic of “\textit{density balanced}” partitions than the classical alternatives.

\subsection{Resource-allocation fairness under extreme population imbalance}

To demonstrate the diagnostic power of the Boltzmann–Shannon Index in settings of extreme socio-economic disparity, we construct a three-community example with highly skewed population shares that reflect real-world developing nations or wealth-concentration scenarios:

\begin{equation}
\textbf{p}_{\text{pop}} = (0.950,\; 0.049,\; 0.001).
\end{equation}

Similar to Subsec.\nobreakspace\ref{sec:two-cluster}, we introduce a single fairness parameter $\beta \in [-1,1]$ that continuously interpolates between perfect proportionality and perfect inversion by defining a resource distribution $r(\beta)$ (which will be used to generate geometric probability distributions $\textbf{q}$ for each value of $\beta$) as the convex combination
\begin{eqnarray}
r(\beta) 
=& \frac{1+\beta}{2}\,\textbf{p}_{\text{pop}} 
  + \frac{1-\beta}{2}\,\textbf{p}_{\text{rev}} \notag \\
=& \frac{1+\beta}{2}
    \begin{pmatrix} 0.950 \\ 0.049 \\ 0.001 \end{pmatrix}
  + \frac{1-\beta}{2}
    \begin{pmatrix} 0.001 \\ 0.049 \\ 0.950 \end{pmatrix},
\label{eq:extreme-resource}
\end{eqnarray}
where $\textbf{p}_{\text{rev}}$ simply reverses the population order. Consequently, $\beta = +1$ represents resources being allocated exactly proportionally to population (total fairness); $\beta =  0$ yields uncorrelated random allocation; and $\beta = -1$ gives almost the entire resource pool ($95 \%$) to the tiniest community of $0.1\%$ of the population (extreme inequality).

For a range of $\beta$ values, we generate $N=500,000$ points (two-dimensions) sampled from a Gaussian distribution, with coordinates representing the resource allocation parameters. Each dimension of the resource space corresponds to a distinct type of resource. The label for each point is assigned according to the population probability distribution $p_{\text {pop }}$.

To ensure that the geometric structure of the resources scales appropriately in accordance with Eq. \ref{eq:extreme-resource}, the allocation vector $r$ is computed as a function of $\beta$. For each $\beta$ value, the coordinates of the points are rescaled as $X_i=U_i R_i V_i^T$, where $U_i$ and $V_i$ are unitary matrices obtained from the singular value decomposition (SVD) of the group (community) matrix $X_i$, and $R_i$ is a diagonal matrix with the allocation parameters $r_i(\beta)$ (corresponding to the $ \mathrm{i}^{th}$ group) along the diagonal.

\begin{figure}[tb]
\centering
\centering
\includegraphics[scale=0.4]{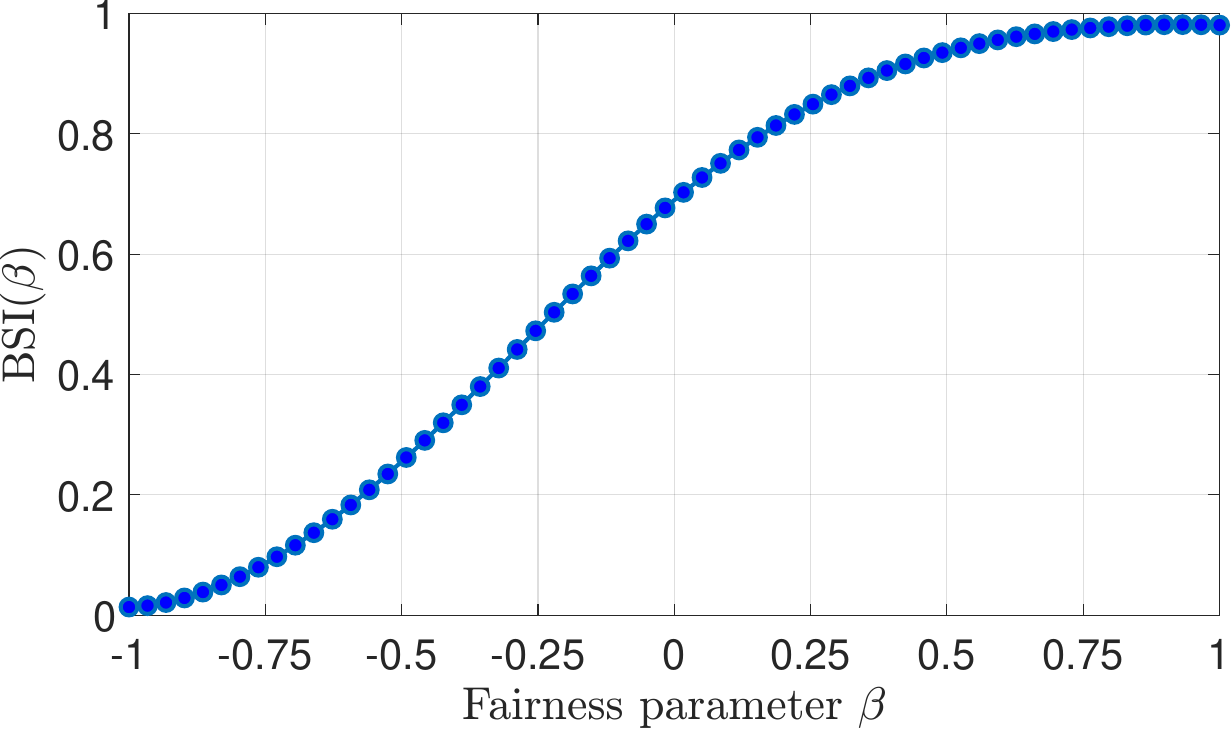}
\caption{\textit{Boltzmann–Shannon Index as a function of the fairness parameter $\beta$ for population shares 95.0 \%,4.9 \%,0.1 \%}. The index is maximal ($\approx 0.98$) under strictly proportional allocations ($\beta = +1$) and collapses to nearly zero when resources are inverted toward the smallest community ($\beta = -1$).}
\label{fig:extreme-resource}
\end{figure}

The resulting relationship (see Figure \ref{fig:extreme-resource}) exhibits a smooth and highly informative curve linking the fairness parameter $\beta$ to the Boltzmann-Shannon Index.  

At complete inversion ($\beta = -1$), when nearly all resources are allocated to the tiniest community ($0.1\%$ of the population), the index collapses to $\operatorname{BSI}(-1)\approx 0.06$. As $\beta$ increases toward neutrality, the BSI gradually increases to approximately $0.70$ at $\beta = 0$ (an interesting value for uncorrelated allocation), and, eventually, reaches $\operatorname{BSI}(1)\approx 0.98$ only under strict proportionality, reflecting the near-perfect alignment between demographic weight and geometric spread.  
The absence of abrupt transitions is itself instructive: in the presence of extreme demographic skew $(95 \%,4.9 \%,0.1 \%)$, truly catastrophic inequality is required to drive the index to near-zero, yet even modest departures from perfect proportionality are consistently and visibly penalized, producing a smooth but uncompromising gradient of fairness.  Of particular interest is the relatively more positive (greater than $0.5$) value of uncorrelated resource allocation.

This smooth behavior confers two immediate practical advantages for quantifying inequality in resource allocation. First, the index produces a continuous and interpretable scalar value in $[0,1]$ that can be reported alongside other established measures, such as the Gini coefficient \cite{gini1912variabilita}, Theil index \cite{theil1967economics}, and Atkinson index \cite{atkinson1970measurement}.  At the same time, it captures a complementary notion of fairness grounded jointly in frequency (the number of members sharing the portion of resources) and geometry (the relative distribution of the underlying allocated resources).  Moreover, unlike purely frequency-based metrics, the BSI accounts simultaneously for both demographic weight and the effective opportunity space available to each group, making it particularly sensitive to scenarios in which a small elite occupies a disproportionately large region of the state space while the majority is compressed into a narrow, resource-poor domain. 

Second, and more importantly, the Boltzmann–Shannon Index is effectively differentiable with respect to the allocation vector $r$, since both the frequency distribution $\textbf{p}$ and the SVD-based geometric distribution $\textbf{q}$ vary smoothly with $r$ (at least numerically). This property enables the BSI to be incorporated directly as a regularization term in constrained optimization formulations for resource allocation. Such formulations naturally arise in public-policy design (health-care budgets, education funding), disaster-relief distribution, and humanitarian aid allocation. Because maximizing the BSI is equivalent to minimizing the Jensen–Shannon divergence between demographic and resource distributions, the resulting optima favor allocations in which each community receives resources in proportion to both its population and its effective spread in outcome space. Although this may not always be the goal of optimization, it can provide a useful regularization term.

In addition, maximizing the Boltzmann–Shannon Index is not universally desirable. In certain domains, a deliberately low BSI may be preferable. A prominent example is the analysis of carbon-emissions across entities (countries, firms, or individuals). In this context, the optimal outcome may be one in which the overwhelming majority of emitters produce very little carbon, and a small number of high-emission activities are those that can be managed or regulated.  Thus, depending on the normative goal, e.g., equity in opportunity versus concentration of harmful outcomes, the BSI can be either maximized (as a fairness regularizer) or minimized (as a harm reduction regularizer).

\section{Conclusion}

In this paper, we introduced the Boltzmann–Shannon Index as a simple but effective way to quantify how well a clustering reflects both the frequency and the geometry of data in continuous spaces. By pairing the distribution of frequencies in cluster labels with a geometric distribution derived from SVD-based volume estimates, the index provides a direct measure of how evenly the state space is partitioned. When these two distributions agree, the index approaches its maximum value (i.e., 1), while when they diverge, the index drops smoothly and predictably towards the lowest possible value (i.e., 0). This behavior gives an immediate sense of whether a partition is “density-balanced” in a meaningful geometric sense, something that traditional clustering indices are not designed to detect.

Through a series of examples, we showed that the Boltzmann–Shannon Index yields a coherent picture across a wide range of settings. In all cases, the index responds precisely to the interactions between population and geometric spread that motivated its construction. Classical measures, by contrast, often give mixed or incomplete signals, either because they depend only on distances or because they ignore the underlying geometry altogether.

Finally, the numerical differentiability of the BSI suggests a practical role beyond diagnostic use. For instance, it can be incorporated directly into optimization frameworks where fairness or balanced representation is a design objective, providing a smooth penalty that favors partitions aligned with both demographic weight and geometric extent. We expect this dual perspective of considering frequency and geometry together to be increasingly valuable in applications involving mixed data, coarse-grained dynamical systems, and modern resource-allocation problems.

\section*{Data Availability}
Data and relevant code for this research work are freely available at: https://github.com/almomaa/Boltzmann\_Shannon\_Index
\bibliographystyle{unsrt}
\bibliography{biblio}

\end{document}